\documentclass[twocolumn,showpacs,amssymb,preprintnumbers,prl]{revtex4}
\usepackage{rotating,epsfig,dcolumn}
\usepackage{graphicx,bm}
\begin{document}
\title{Three-body monopole corrections to the realistic interactions}
\author{ A. P.~Zuker$^a$}
\affiliation{
(a) IReS, B\^at27, IN2P3-CNRS/Universit\'e Louis
Pasteur BP 28, F-67037 Strasbourg Cedex 2, France}
\date{\today}
\begin{abstract}
  It is shown that a very simple three-body monopole term can solve
  practically all the spectroscopic problems---in the $p$, $sd$ and
  $pf$ shells---that were hitherto assumed to need drastic revisions
  of the realistic potentials.
\end{abstract}
\maketitle 

The first exact Green's function Monte Carlo (GFMC) solutions for
$A>4$ nuclei confirmed that two-body (2b) interactions fell short of
perfectly reproducing experimental data~\cite{pud97}. The inclusion of
a three-body (3b) force lead to excellent spectroscopy, but some
problems remained for the binding and symmetry energies and spin orbit
splittings. No core shell model calculations (NCSM)~\cite{zhe93} have
recently developped to the point of approximating the exact solutions
with sufficient accuracy to provide a very important---though
apparently negative---result in $^{10}$B~\cite{nav02}: While in the
lighter systems the {\em spectra} given by a strict two-body potential
are not always good---but always acceptable---in $^{10}$B, the
spectrum is simply very bad.

My purpose is to show the striking analogy between this situation and
what occurs in conventional ($0\hbar \omega$) shell model calculations
with realistic G-matrices, and then explain how a very simple 3b term
can solve practically all the spectroscopic problems---in the $p$,
$sd$ and $pf$ shells---that were hitherto assumed to need drastic
revisions of the realistic (R) 2b potentials.  

The first realistic matrix elements~\cite{kuo66}, and the first large
scale shell model codes~\cite{fre69} appeared almost simultaneously.
Calculations for up to five particles in the $sd$ shell gave very
satisfactory results, but the spectrum of $^{22}$Na (i.e., $(sd)^6\,
T=0$) was very bad~\cite{hal71}. (Note that $^{10}$B is $(p)^6\, T=0$).
At the time nobody thought of 3b forces, and naturally the blame was
put on the 2b matrix elements ($V_{stuv}^{JT},\, stuv$ are subshells).
The proposed phenomenological cures amounted to fit them to the
experimental levels. Two ``schools'' emerged: One proposed to fit them
all (63 in the $sd$ shell), and lead eventually to the famous USD
interaction~\cite{wil84,bro88}. The alternative was to fit only the
centroids, given in Eqs.~(\ref{eq:v},\ref{eq:ab}).
\begin{eqnarray}
  \label{eq:v}
  V_{st}^T&=&\frac{\sum_J
  V_{stst}^{JT}(2J+1)[1-(-)^{J+T}]\, \delta_{st}}{(2J+1)[1-(-)^{J+T}]
 \,  \delta_{st}},\\ 
\label{eq:ab}
a_{st}&=&\frac{1}{4}(3V_{st}^1+V_{st}^0), \hspace{10pt}
b_{st}=V_{st}^1-V_{st}^0\\
  \label{eq:nrs}
  n_{st}&=&\frac{1}{1+\delta_{st}}n_r(n_s-\delta_{st}), \\
\label{eq:trs}
  T_{st}&=&\frac{1}{1+\delta_{st}}(T_r\cdot
  T_s-\frac{3}{4}n_{st}\, \delta_{st})\\ 
\label{eq:hm}
 H_m&=&\sum_s \varepsilon_s\, n_s+\sum_{s\le t} (a_{st}\,
 n_{st}+b_{st}\, T_{st})
\end{eqnarray}
They are associated to the 2b quadratics in number ($n_s$) and isospin
operators ($T_s$), Eqs.~(\ref{eq:nrs},\ref{eq:trs}), and they define the
monopole Hamiltonian, Eq.~(\ref{eq:hm}), in which we have added the
single particle (1b) term. The idea originated in Ref.~\cite{pas76},
where it was found that the Kuo Brown (KB) interaction in the $pf$
shell~\cite{kuo68} could yield excellent spectroscopy through the
modifications ($f\equiv f_{7/2},\, r\equiv f_{5/2},\, p_{3/2},\, p_{1/2}$),
\begin{equation}
  \begin{array}{l}
V_{fr}^T(\text{KB1})=V_{fr}^T(\text{KB})-(-)^T\,300 \text{ keV},\\
V_{ff}^0(\text{KB1}) = V_{ff}^0(\text{KB})-350 \text{ keV},\\
V_{ff}^1(\text{KB1})= V_{ff}^1(\text{KB})-110 \text{ keV}.
  \end{array}
  \label{eq:kb1}
\end{equation}
The validity of this prescription was checked in perturbative
calculations~\cite{pov81}, and convincingly confirmed for $A=47-52$
once exact diagonalizations became
feasible~\cite{cau89,zuk97,pov01a,pov01b,note1}.

In what follows $f$ will stand generically for ($p_{3/2},\, d_{5/2},
\, f_{7/2}$) in the ($p,\, sd,\, pf$) shells respectively. Obviously
$r= p_{1/2}$ and $r\equiv d_{3/2},s_{1/2}$ for the $p$ and $sd$ shells. 

Nowadays the 2b $NN$ potentials are nearly perfect, and the
calculations are exact. Therefore, the blame for bad spectroscopy must
be put on the absence of 3b terms. Which means that the monopole
corrections {\em must} be 3b and Eq.~(\ref{eq:hm}) must be
supplemented by
\[\sum_{s\, t\, u} (a_{stu}\, n_{stu}+b_{stu}\, T_{stu}),\] 
where $n_{stu}\equiv n_{rs}n_t$, or $n_r(n_r-1)(n_r-2)/6$ and similar
forms for $T_{stu}$. To simplify matters we---tentatively---allow only
contributions of the type $n_{st}(n-2)$ and $T_{st}(n-2)$, i.e., 2b
terms modulated by the total number of particles $n$. It should be
borne in mind that a 3b interaction also produces 2b pieces in the
model space, exactly in the same way that the 2b interaction produces
the single particle splittings by summing over the core orbits $c$ of
degeneracy $D_c$,
\begin{equation}
  \label{eq:eps}
\sum_c a_{sc}n_s\, n_c=n_s\sum_c a_{sc}\, D_c\equiv n_s \varepsilon_s.  
\end{equation}
Note that a 3b potential will produce both 1b and 2b terms. We need
not worry about the former because they correct $\varepsilon_s$ {\em
  which will be taken from experiment as traditionally done}. The
latter, together with the 3b part will transform the realistic (R) 2b
centroid $V_{st}^T(\text{R})$ into
$V_{st}^T(\text{R1})=V_{st}^T(\text{R})+(\alpha^T_{st}+\beta^T_{st}
n)\equiv V_{st}^T(\text{R})+\chi'^{\, T}_{st}$.  

$H_m$ can be characterized
by demanding correct single particle and single hole spectra around
closed shell nuclei~\cite{duf99}. This set ($cs\pm 1$) is taken to
include the differences in binding energies (gaps)
$2BE(cs)-BE(cs+1)-BE(cs-1)$.  The major monopole correction involves
the gaps around $^{12}$C, $^{28}$Si, $^{48}$Ca and $^{56}$Ni which are
too small to produce the observed double magicity~\cite{note2}. It
will be taken care by a single linear form $\kappa\equiv \kappa(n)$.
The generalization of Eq.~(\ref{eq:kb1}) is then

\begin{equation}
  \begin{array}{l}
V_{fr}^T(\text{R1})=V_{fr}^T(\text{R})-(-)^T\, \kappa+\chi^T_{fr}\, ,\\
V_{ff}^T(\text{R1}) = (V_{ff}^T(\text{R})-1.5\, \kappa)\, 
\delta_{T0}+\chi^T_{ff}\, ,\\  
V_{rr'}^T(\text{R1})= V_{rr'}^T(\text{R})+\chi^T_{rr'}. 
  \end{array}
  \label{eq:R1}
\end{equation}
The single particle splittings above the $f$ closures are quite well
given by some R interactions. Hence the corrective term
$\chi^T_{fr}$---which will prove useful in the $sd$ shell---is most
likely to have a 2b origin. $\chi^T_{ff}$ is introduced only for
completeness and will be altogether disregarded.  $\chi^T_{rr'}$ must
play an important role because the single hole states (at $A=15$, 39
and 79)~\cite{note3} are severely missed. However, they have little
influence on the nuclei we shall study (at the beginning of the
shells).

\begin{figure}[t]
  \begin{center}
    \leavevmode 
    \epsfig{file=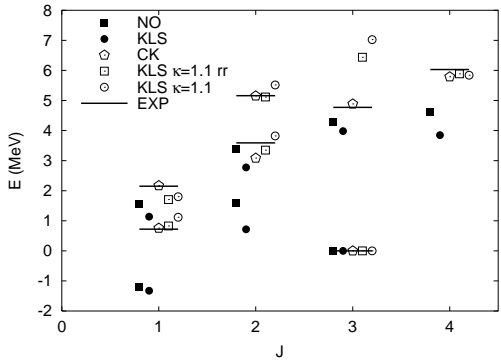,width=7cm}
    \caption{Excitation energies for $^{10}$B referred to the $J=3$
    lowest state. See text} 
    \label{fig:10}
  \end{center}
\end{figure}

For the $^{10}$B spectrum in Fig.~\ref{fig:10} the black squares show
the results of N\'avratil and Ormand (NO)~\cite[Fig 4,
6$\hbar\Omega$]{nav02} for the low lying $T=0$ states in $^{10}$B. The
black circles correspond to the bare KLS G-matrices~\cite{kah69,lee69}
used in \cite{abz91}.  The agreement with experiment
(lines)~\cite{ajz88} is poor {\em but the agreement between the
  calculations is good}. This is not a joke, but an important remark:
NO provides the {\em foundation} for a conventional G-matrix study. As
emphasized over the years~\cite{pas76,abz91,duf96}, the realistic
G-matrices are very close to one another and will provide good spectra
once monopole corrected. Absolute energies and strength functions are
another matter, and much remains to be learnt from exact and no-core
results.

The open pentagons in Fig.~\ref{fig:10} correspond to the classic
Cohen Kurath fit~\cite{coh65} (CK). The open squares and circles refer
to the KLS interaction with a $\kappa=1.1$ correction in
Eq.~(\ref{eq:R1}). The open squares test the influence of the
$\chi^T_{rr}$ term through a uniform attraction of 1.5 MeV (in CK it
is about 3 times as large). Conclusion: there is not much to choose
between the two LKS corrected cases. Moreover, they are practically as
good as CK except for the second $J=3$ level.

There are two reasons not to dwell any longer in the $p$ shell. The
first is that the aim of this letter is to show that the monopole
corrections must be 3b, i.e., $\kappa$ must be linear in $n$, which
demands examining cases of sufficiently different $n$. Unfortunately
this is impossible without bringing in the other possible
contributions: For example, $\chi^T_{rr}$ is not very significant in
$^{10}$B ($n=6$), but it is important in $^{12}$C ($n=8$) and crucial
in $^{14}$N ($n=10$). Therefore, there is no way of exploring what a
single term in Eq.~(\ref{eq:R1}) does: all must contribute. As it
happens---and this is the second reason---the full exploration {\em
  has been done}~\cite{abz91}, and the results were excellent. At the
time, the problem was that the 3b contributions turned out to be large
and important, and the authors did not know what to do with them.
\begin{figure}[t]
  \begin{center}
    \leavevmode 
    \epsfig{file=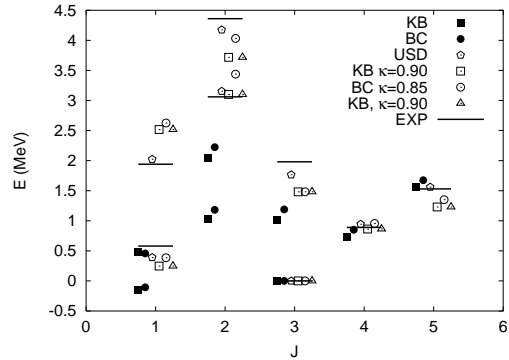,width=7cm}
    \caption{Excitation energies for $^{22}$Na referred to the $J=3$
    lowest state. See text.}
    \label{fig:22}
  \end{center}
\end{figure}

For the $^{22}$Na spectrum in Fig.~\ref{fig:22} the black squares show
the results for the venerable KB~\cite{kuo66}. The black circles
correspond to the BonnC (BC) G-matrices~\cite{hjo95,note4}. The
agreement with experiment (lines)~\cite{end90} is poor {\em but the
  agreement between the calculations is good}. Again, this is not a
joke, but an important remark: as mentioned, there are very little
differences between the realistic G-matrices. The open pentagons
correspond to Wildenthal's USD~\cite{wil84}. The open squares and
circles refer to the KB and BC interactions with $\kappa=0.9$ and 0.85
corrections respectively. We shall come to the triangles soon.  Though
USD is closer to experiment, the corrected R interactions do
definitely well.

\begin{figure}[t]
  \begin{center}
    \leavevmode 
    \epsfig{file=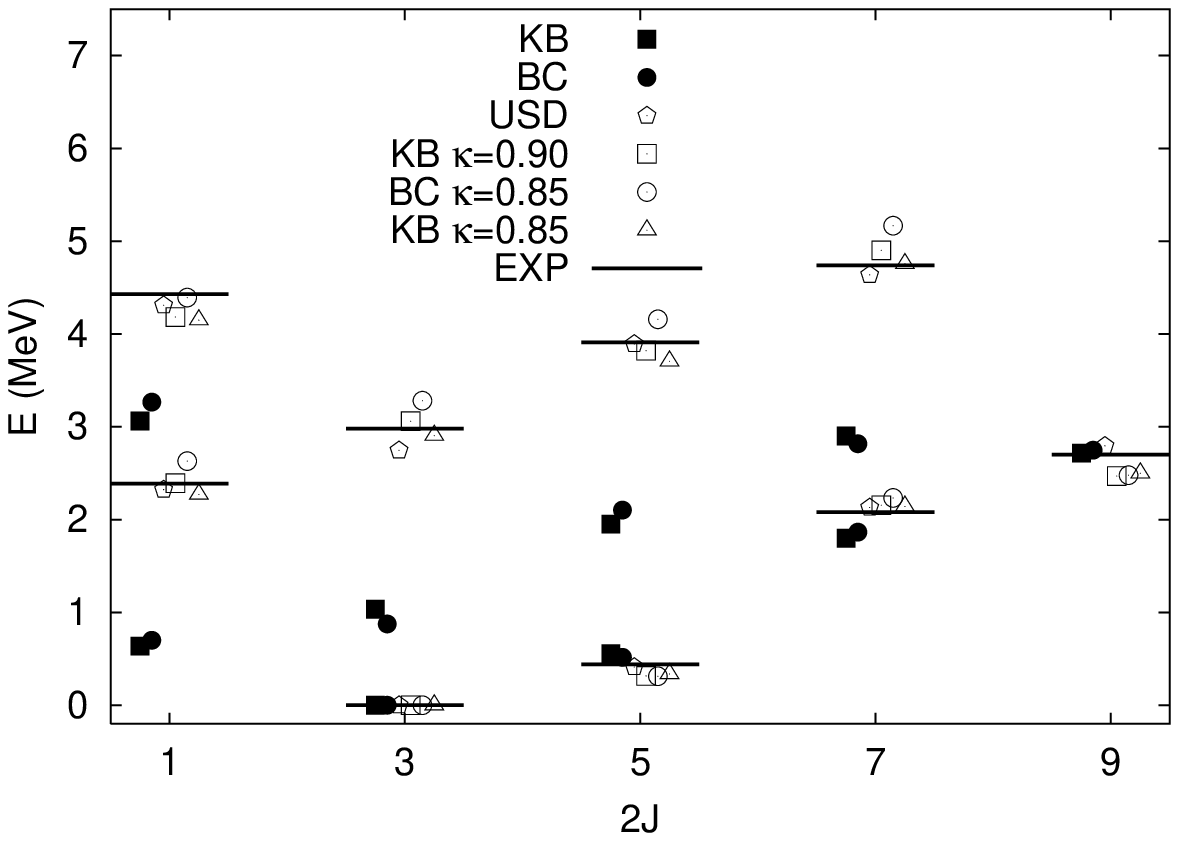,width=7cm}
    \epsfig{file=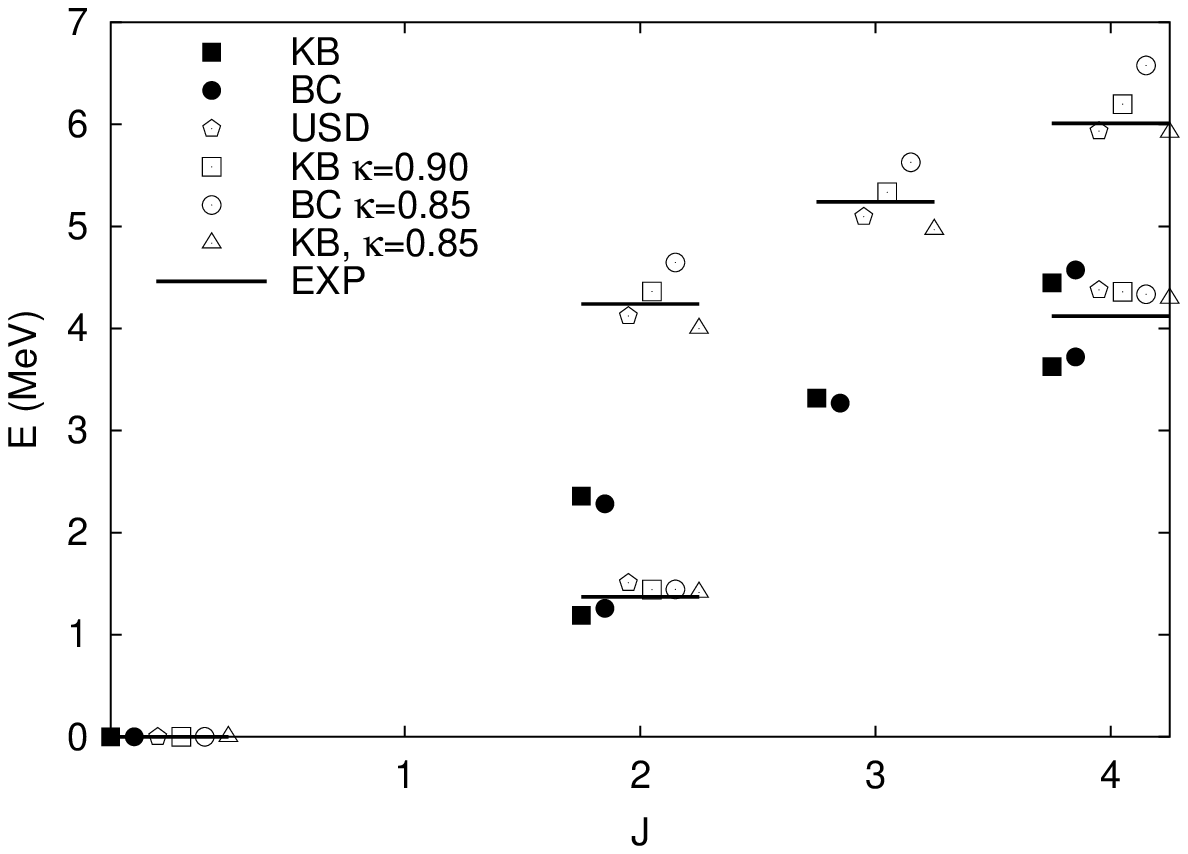,width=7cm}
    \caption{Excitation energies for $^{23}$Na and $^{24}$Mg referred
      to the $J=3/2$ and 0 lowest states repectively. See text.}
    \label{fig:2324}
  \end{center}
\end{figure}

The story repeats itself for $^{23}$Na and $^{24}$Mg in
Fig.~\ref{fig:2324}. The notations are the same as in
Fig.~\ref{fig:22}. The agreement with experiment is now truly
satisfactory, and the plotting technique adopted makes the physics
quite evident: the trouble with a 2b-only description is that the
excited band $K=1/2$ in $^{23}$Na, and the $K=2\, (\gamma)$ band in
$^{24}$Mg are too low.

\begin{figure}[t]
  \begin{center}
    \leavevmode  
    \epsfig{file=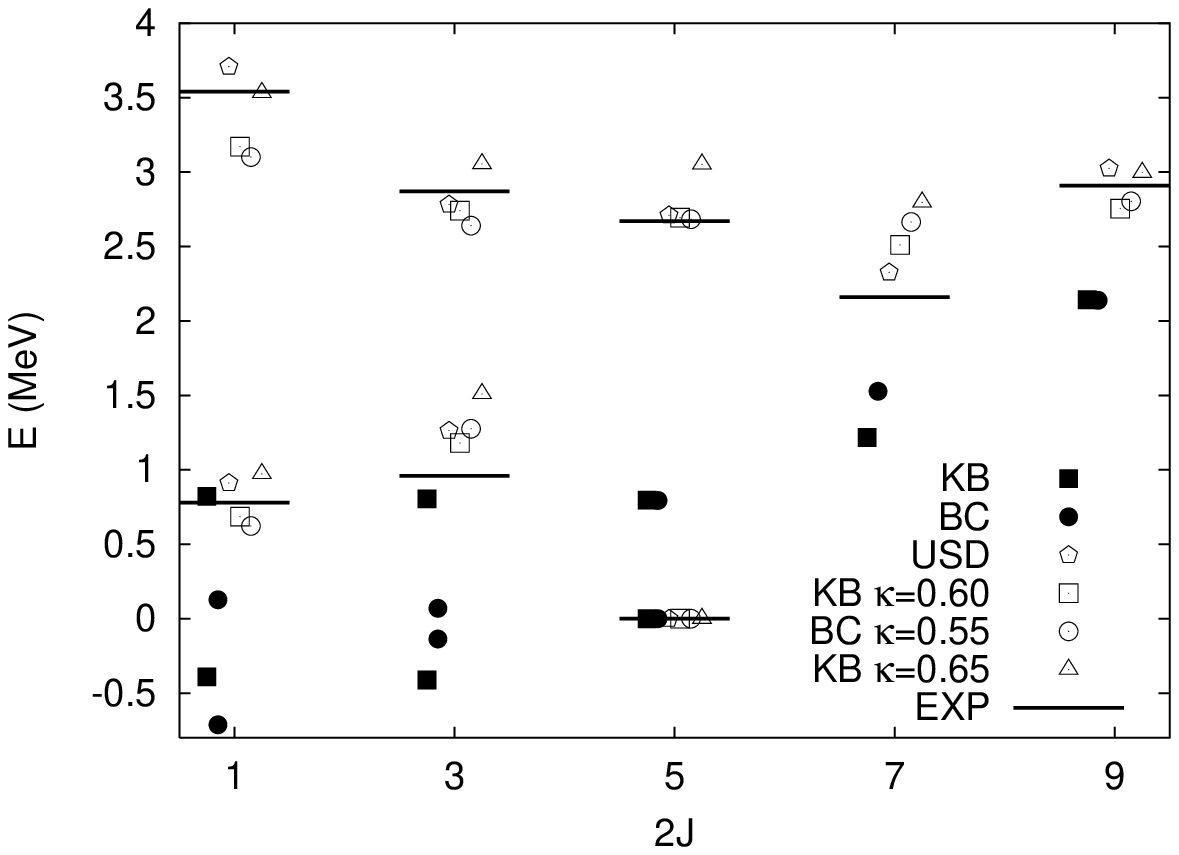,width=7cm}
    \epsfig{file=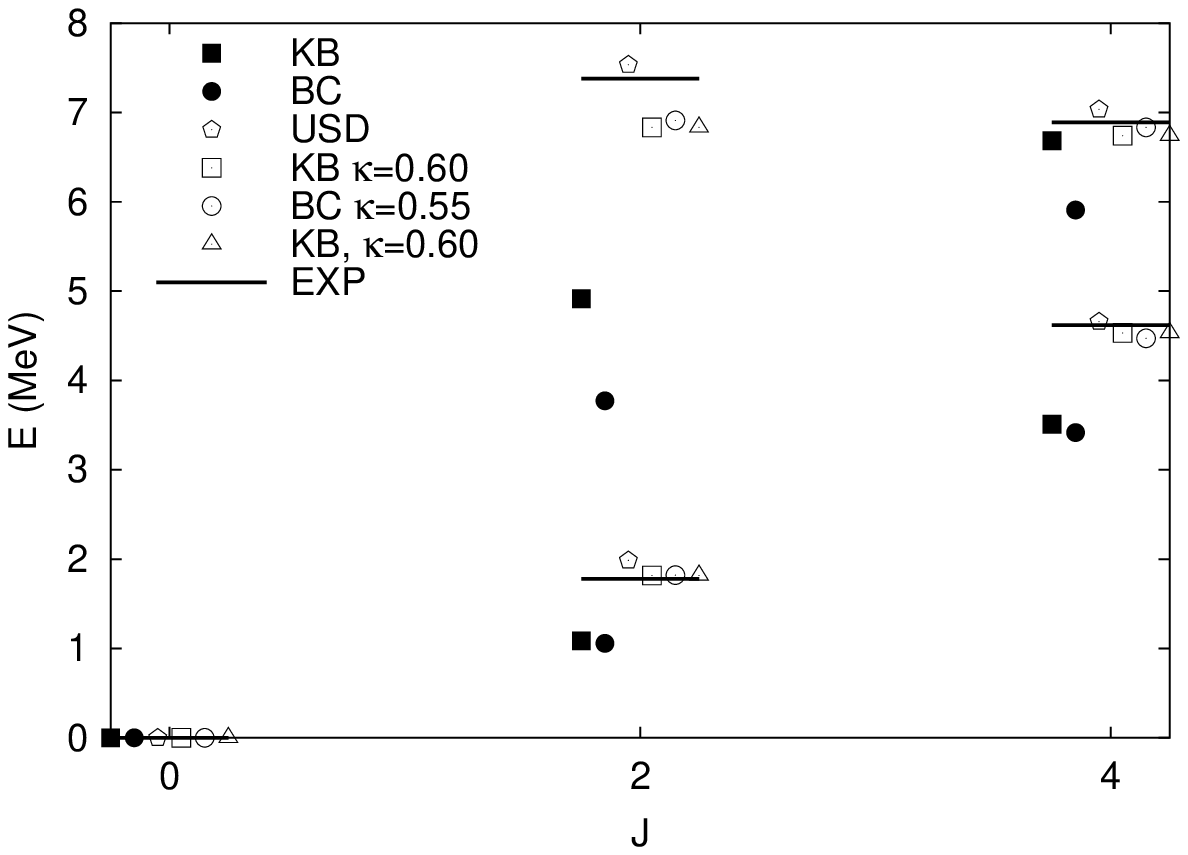,width=7cm}
    \epsfig{file=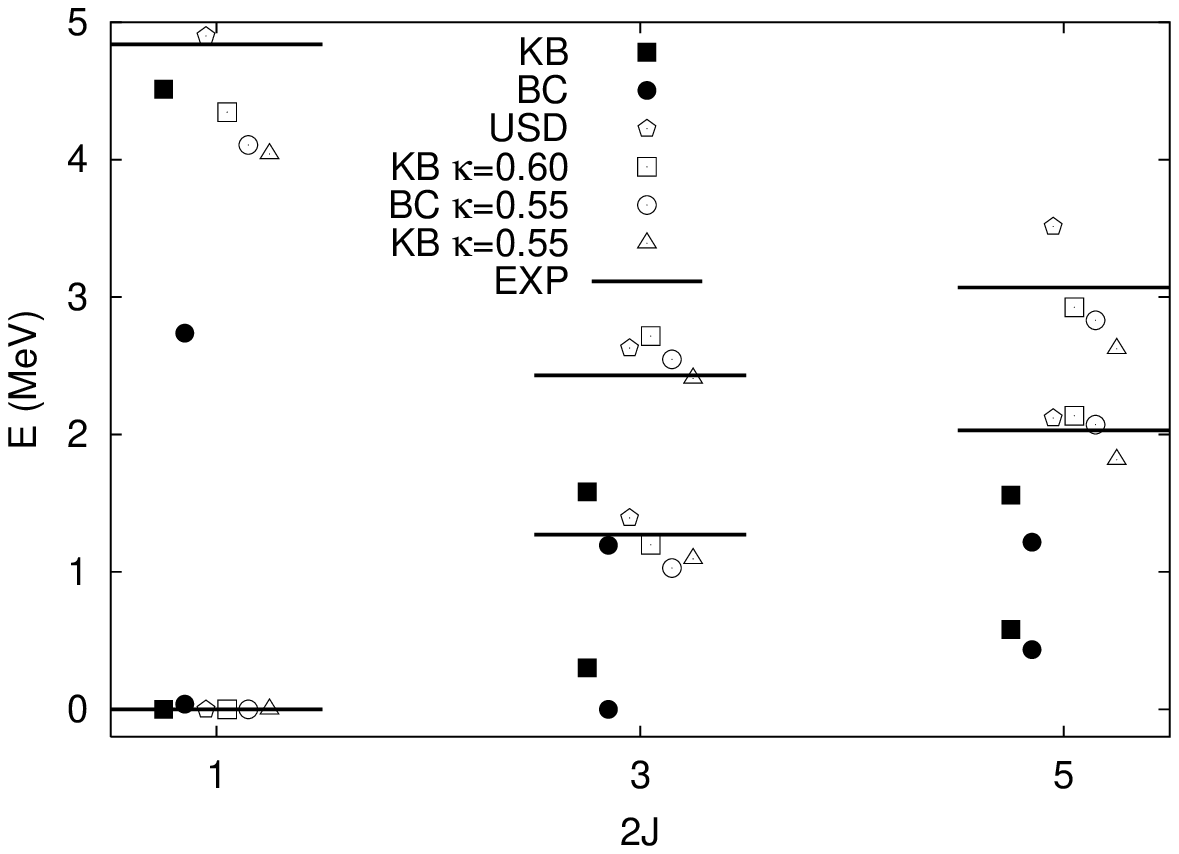,width=7cm}
    \caption{Excitation energies for $^{27}$Si, $^{28}$Si and,
    $^{29}$Si  referred to the $J=5/2$, 0 and 1/2 
    lowest states respectively. See text.}
    \label{fig:2729}
  \end{center}
\end{figure}

The open triangles in Figs.~\ref{fig:22} and~\ref{fig:2324} show what
happens with KB when---instead of keeping $\kappa$ fixed---we increase
it by steps of 0.5 per $n$. Though there is an improvement, it is not
sufficient to claim the irrefutability of a 3b mechanism. The proof
comes when we move to Figs.~\ref{fig:2729}: In $^{27}$Si, $^{28}$Si,
and $^{29}$Si the local value of $\kappa$ (open squares and circles)
has decreased to 0.60 for KB and to 0.55 for BC. {\em A constant
  $\kappa$ is totally ruled out}, while he linear law (triangles) does
quite well. Clearly, the 3b terms are indispensable. The superb
2b-only USD fit was obtained mostly through the massacre of a strong
$JT=20$ pairing term that is a constant feature of the R interactions,
which makes USD R-incompatible~\cite[Section V]{duf96}. This has been
known for some time and it is only occasionally that trouble may
arise. The problem has been the difficulty, so far, of obtaining an
R-compatible fit of comparable quality. The mild exception comes
from~\cite{abz91} where, as in the $p$ shell, the 3b contributions
turned out to be so large and important, that the authors did not know
what to do with them.

\begin{figure}[t]
  \begin{center}
    \leavevmode 
    \epsfig{file=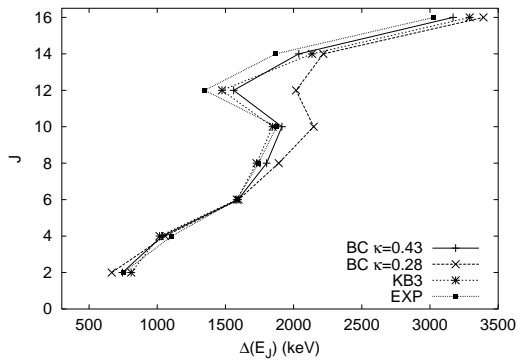,width=7cm}
    \caption{Backbending in $^{48}$Cr. See text.}
    \label{fig:bb}
  \end{center}
\end{figure}

In the $pf$ shell KB1 (or KB3) is very good for $A=47$-52 but it
produces too large a gap at $^{56}$Ni (7.5 MeV against the observed
6.3 MeV). The most serious problem comes from the first
$BE2(2\Longrightarrow 0)$ transition in $^{58}$Ni which falls short of
the observed value (140 e$^2$fm$^4$) by a factor $\approx 0.4$. Here
it is expedient to replace the constants in Eq.~(\ref{eq:kb1}) by
linear terms that have the same value at $A=48$, and are reduced by a
factor $\approx 0.7$ at $A=56$ ( Eq.~(\ref{eq:R1}) works as well).
The situation in $^{56}$Ni becomes consistent with experiment but in
$^{58}$Ni it remains unacceptable.  The problem is solved by BonnC
(BC)~\cite{hjo95}, with the same  $\kappa$ reduction of
$\approx 0.7$ in going from $A=48$ to $A=56$. The key difference
between KB anb BC is that the intensity of the quadrupole force (in
MeV, extracted as in~\cite{duf96}) is 2.7 for KB and 3.2 for BC. This
discrepancy is somewhat disturbing, but it does alter the basic fact
that 3b monopole terms are necessary. Fig.~\ref{fig:bb} shows that for
$\kappa=0.43$, BC produces a backbening pattern in $^{48}$Cr that is
practically as good as the KB3 one. At $\kappa=0.28$---the correct
value around $A=56$---the agreement with experiment is destroyed.

There are several other indications that a 3b interaction is
essential.  Perhaps the most significant is the following: The
monopole centroids $V^T_{f_{7/2}(sd)}$ must be such that when
$f_{7/2}$ fills the $d,\, (l=2)$ orbits are depressed with respect to
the $s, \, (l=0)$ one~\cite{duf99}. {\em However}, it is clear from the
spectrum and the spectroscopic factors in $^{29}$Si that the filling of
$d_{5/2}$ favours the $p, \, (l=1)$ orbit(s) over the $f,\, (l=3)$
ones~\cite{end90}. A 2b-only assumption leads to a contradiction: if
$f_{7/2}$ acting on the $sd$ shell favours the larger $l$ orbits,
$d_{5/2}$ acting on the $pf$ shell must do the same. Without
unacceptable ad-hoc assumptions, a 2b mechanism
cannot do otherwise but a 3b one can.

From what we have seen, 3b {\em monopole} forces make things simpler,
and there are good reasons to believe that the formidable task of a
full treatment---including multipole terms---need not be inevitable .
A recent generation of 3b potentials~\cite{pie01} has made it possible
for the exact solutions to eliminate the more offending aspects of the
2b $^{10}$B spectrum~\cite{pie02}.  It will be of much interest to
check whether the underlying mechanism corresponds to the one proposed
in this letter.  At any rate, a full characterization of the 3b
potentials is not an easy matter, and it could be hoped that
information coming from shell model studies may prove valuable.
Especially at a time when GFMC and no-core calculations have rigorously
established the basic reliability of such studies.

Several obsevations of Alfredo Poves and Fr\'ed\'erique Nowacki have been of
great help.


\begin{thebibliography}{70}
\bibitem{note1} For ``historical'' reasons, these calculations use the
  KB3 interaction~\cite{pov81}, which differs from KB1 in some innocuous non
  monopole changes.
\bibitem{note2} The closed shell component in the ground state of
  these nuclei may be as small as 30\% in $^{28}$Si but the large gaps
  and the spectroscopic factors for the $cs\pm 1$ states indicate
  strong magicity.
\bibitem{note3} $^{80}$Zr is not a closure, but we can always
  calculate the monoplole properties around a closed $pf$ shell.
\end{thebibliography}
\end{document}